\documentclass[12pt,english,a4paper]{iopart}
\usepackage[T1]{fontenc}
\usepackage[latin1]{inputenc}
\usepackage{iopams}
\usepackage{babel}
\usepackage{graphicx}
\usepackage{multirow}
\usepackage{booktabs}

\begin{document}

\title{Physisorption of an organometallic platinum complex on silica. An {\it ab initio} study}

\author{Juan Shen, Kaliappan Muthukumar, Harald O. Jeschke, Roser Valent{\'\i}}

\address{Institut f{\"u}r Theoretische Physik, Goethe-Universit{\"a}t Frankfurt,
Max-von-Laue-Str. 1, 60438 Frankfurt, Germany}

\begin{abstract}
  The interaction of trimethyl methylcyclopentadienyl platinum
  (MeCpPtMe$_3$) with a fully hydroxylated SiO$_2$ surface has been
  explored by means of {\it ab initio} calculations.  A large slab
  model cut out from the hydroxylated $\beta$-cristobalite
  SiO$_2$ (111) surface was chosen to simulate a silica surface.
  Density functional theory  calculations were performed to
  evaluate the energies of MeCpPtMe$_3$ adsorption to the SiO$_2$
  surface.  Our results show that the physisorption of the molecule is
  dependent on both (i) the orientation of the adsorbate and (ii) the
  adsorption site on the substrate.  The most stable configuration was
  found with the MeCp and Me$_3$ groups of the molecule oriented
  towards the surface. Finally, we observe that van-der-Waals 
  corrections are crucial for the stabilization of the molecule on the
  surface. We discuss the relevance of our results for the growth of
  Pt-based nanostructured materials via deposition processes such as
  electron beam induced deposition.
\end{abstract}


\submitto{\NJP}

\maketitle

\section{Introduction}\label{sec:introduction}

Trimethyl methylcyclopentadienyl platinum (MeCpPtMe$_3$) is an
important organometallic precursor that is widely used to deposit Pt
nanostructures, not only in electron beam induced deposition (EBID),
but also in focused ion beam deposition, chemical vapour deposition,
laser induced chemical processing, and atomic layer deposition
processes~\cite{sens-2010-10-9847,jap-2011-109-063715,jpcc-2009-113-2487,surfS-2011-605-257,jvstb-2008-26-2460,jap-2009-106-074903,surfSR-2010-65-1,jvstb-1990-8-1826,jvstb-2002-20-590,jvstb-1992-10-2695,cvd-2003-9-213,prl-2003-91-066102,cm-1992-4-162,tsf-1997-303-136,tsf-1992-218-80,acbE-2010-101-54}.

For most of the deposition processes mentioned above, there is a lack
of detailed understanding of the  growth mechanism of nanostructures at
the molecular level. While for instance in EBID, several Monte Carlo
simulations have been performed to study the effects of electron
energy, probe size, substrate thickness and deposit composition
considering the primary, secondary and backscattered
electrons~\cite{nanote-2006-17-3832,ultram-2005-103-17,scan-2006-28-311,
  microE-2002-61-693,apl-2003-82-3514,jap-2005-98-084905}, very little
is known about the detailed geometry of adsorption configurations and
the possible dissociation processes upon adsorption. If the various
deposition processes are to evolve from a set of empirical recipes for
producing nanostructures to viable scientific tools for
nanofabrication, a complete understanding of the interactions between
the substrate, adsorbed precursor molecules, and electrons in the case
of EBID, is fundamental and must be
developed~\cite{surfS-2011-605-257,jap-2008-104-081301,chemEJ-2007-13-9164}.
In Ref.~\cite{muthupaper}, a first attempt to investigate the
adsorption of tungsten hexacarbonyl (W(CO)$_6$) precursor molecules
on a SiO$_2$ substrate partly modified due to electron irradiation
was performed by means of density functional theory (DFT) calculations.  In
the present work we consider the physisorption process of the more
complex molecule MeCpPtMe$_3$ on a hydroxylated SiO$_2$ substrate
within DFT with and without inclusion of long-range van der Waals
(vdW) interactions. This study should serve as the starting point for
growth of nanostructures in various deposition processes.  SiO$_2$ is
chosen as substrate in our calculations since it is one of the most
important substrates in the semiconductor industry, and is widely used
for instance in the EBID process.

Physisorption properties have been studied in several systems. For
example, Rimola {\it et al.}~\cite{pccp-2010-12-6357} investigated the
adsorption of benzene and benzene-1,4-diol on hydrophobic and
hydrophilic silica surfaces and found that the adsorption of the
aromatic molecules on the hydrophobic silica surface is dictated by
vdW interactions and the adsorption energies for the hydrophilic
surfaces are almost doubled with respect to the hydrophobic surface
due to H-bonding interactions between the substrate and the
adsorbate. Mian {\it et al.} performed DFT calculations to investigate
the catechol adsorption on silica surfaces and indicated that both the
hydroxyls and phenylene ring of catechol contribute to its strong
adhesion due to hydrogen bonding and vdW
interactions~\cite{jpcc-2010-114-20793}.  Atodiresei {\it et al.}
investigated the adsorption mechanism of a single pyridine molecule on
Cu(110) and Ag(110) surfaces and found that vdW corrections are very
important for the geometry and electronic structure of flat adsorption
configurations~\cite{prb-2008-78-045411,prl-2009-102-136809}.  Our
results for MeCpPtMe$_3$ on SiO$_2$ show that the adsorption of the
molecule is highly sensitive to both (i) the molecule orientation and
(ii) the adsorption site on the substrate and we find that only
certain geometries are favored in the process. Moreover, inclusion of
vdW corrections are crucial for stabilizing the molecule on the
surface.

The paper is organized as follows: in the next section the methods
used for the DFT calculations are introduced. In Section 3 the
electronic properties of the precursor molecule, the substrate as well
as the complex molecule-substrate are presented and discussed with
special emphasis on the importance of the vdW corrections. Further, we discuss the
 relevance of our results for the deposition of nanostructures  and in
Section 4 we provide a summary of our work.

\section{Methods and computational details}\label{sec:methods}

\begin{figure}
\begin{center}
\includegraphics[width=0.5\textwidth]{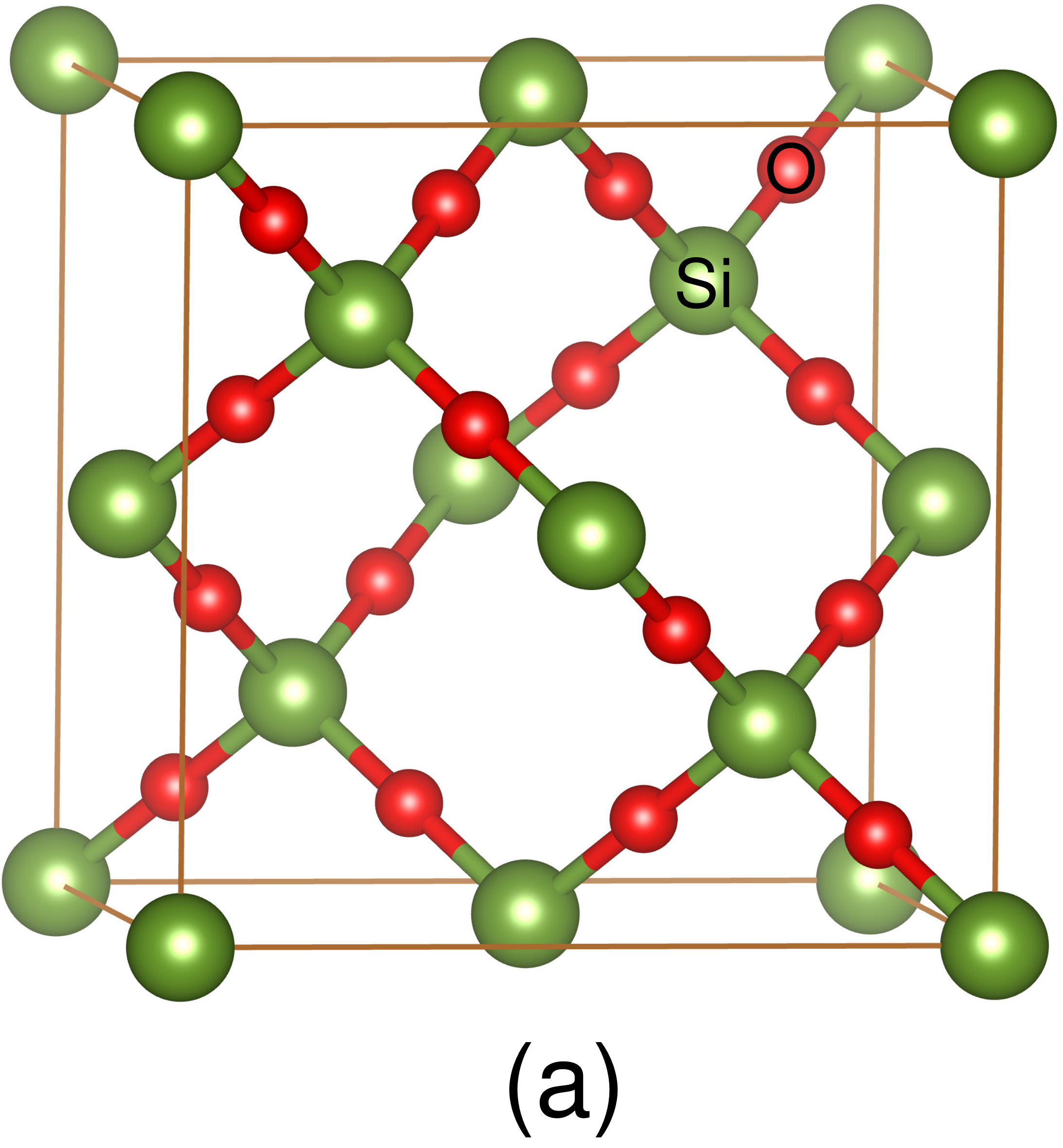}
\includegraphics[width=0.5\textwidth]{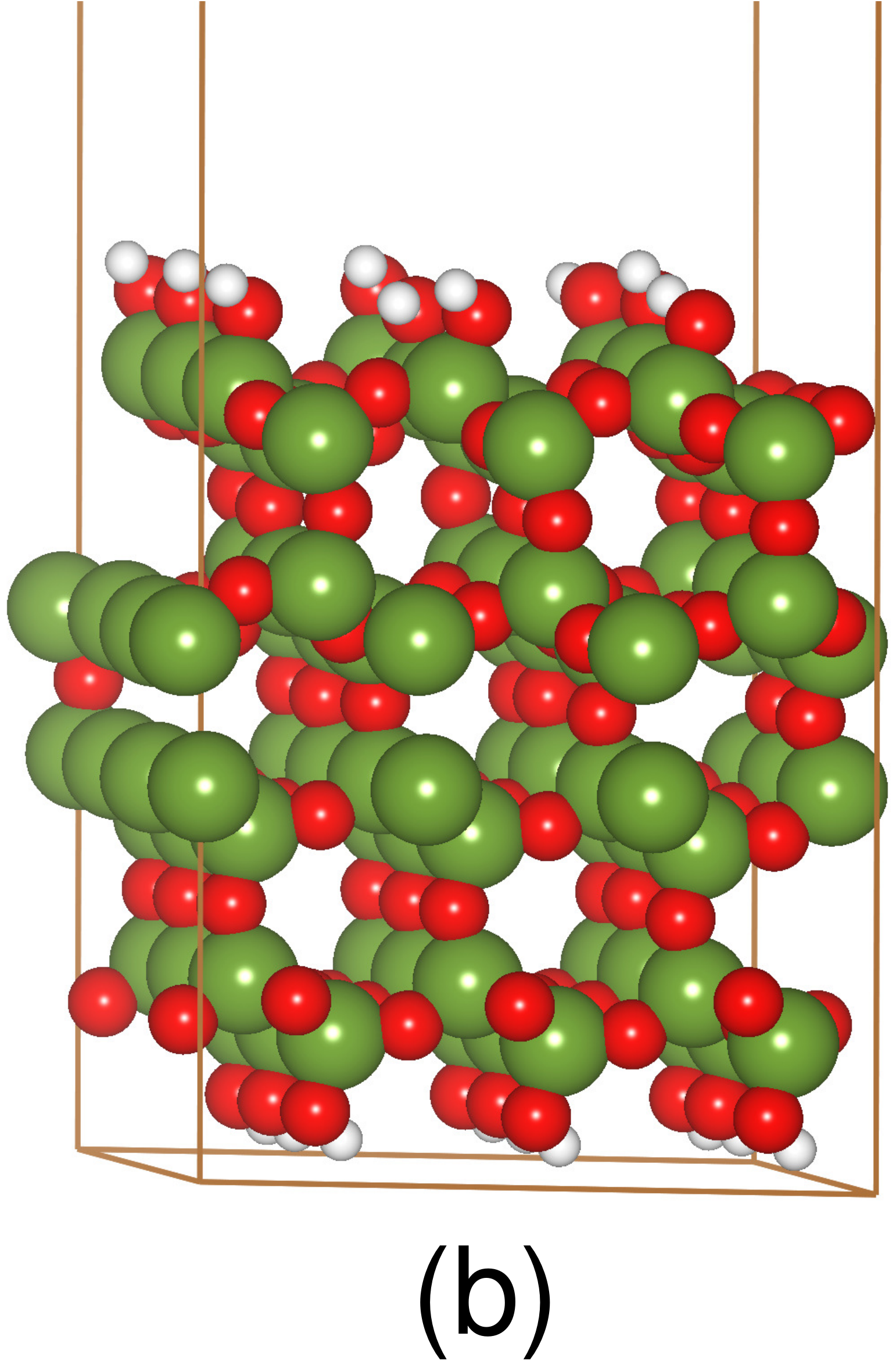}
\end{center}
\caption{(a) Bulk $\beta$-cristobalite SiO$_2$. (b) Optimized structure of the clean surface, cut in [111] crystallographic direction.}
\label{fig:sio2}
\end{figure}

Silicon dioxide (SiO$_2$) is a typical polymorphic material with
structures ranging from amorphous to various types of crystalline
phases. The $\beta$-cristobalite crystalline phase and amorphous
SiO$_2$ have a locally similar structure.  They have similar densities
and refractive indices, and a rather convenient cubic cell.
Therefore, $\beta$-cristobalite has been frequently used as a model
crystal to represent the structure of amorphous
SiO$_2$~\cite{prb-1991-44-11048,prb-2005-72-165410,apl-2008-93-202110,prb-2001-64-195330,jpcb-2005-109-14954}
and will be also considered here as a model substrate.

The $\beta$-cristobalite SiO$_2$ is cubic with eight molecules in the
conventional unit cell (Figure~\ref{fig:sio2}~(a)).  Considering the
large size of the precursor molecule MeCpPtMe$_3$, we need to find a
compromise between computational cost and sufficiently large distances
between adsorbates.  Thus, a four-layer slab model containing 216
atoms was cut from bulk $\beta$-cristobalite (space group $F\,d\bar{3}m$)
with the experimental lattice constant 7.16~{\AA} along the [111]
direction~\cite{muthupaper}. We rearranged the resulting structure by adding OH hydroxyl
groups to the unsaturated Si atoms exposed at the surface to avoid
dangling bond effects. The bottom O atoms resulting from the formation
of the surfaces are also saturated with H atoms.  In fact, a realistic
model for a substrate prepared under nonvacuum conditions is
hydroxylated SiO$_2$.  A vacuum region of approximately 30~{\AA}
without considering the adsorbate was placed between slab layers. The
geometry is shown in Figure~\ref{fig:sio2}(b).

Hydroxyl coverage on silica has been previously measured to be
4.9/nm$^{2}$ OH groups at 180-200 $^\circ$C and 3.6/nm$^{2}$ at
300$^\circ$C, independent of the structural
characteristics~\cite{langm-1987-3-316}.  Our model built by a
crystalline $\beta$-cristobalite [111] facet has a silanol density of
4.5/nm$^2$ close to that observed on amorphous silica and therefore is
a good representation of the experimental environment.

Structure relaxation and electronic properties of the substrate and
the complex molecule-substrate were performed in the framework of
DFT by using the Vienna {\it ab initio}
Simulation Package (VASP)~\cite{prb-1993-47-558,cms-1996-6-15} with a
projector augmented wave (PAW) basis
set~\cite{prb-1994-50-17953,prb-1999-59-1758}.  Exchange and
correlation effects were treated in the generalized gradient
approximation (GGA) in the formulation of Perdew, Burke, and Ernzerhof
~\cite{prl-1996-77-3865}.  The plane wave basis set used to
represent the valence electron density was cut off at 400 eV. The
Brillouin zone was sampled using a $(2 \times 2 \times 1)$
$\Gamma$-centred mesh.  Structural optimizations were performed by
minimizing the forces on all atoms to less than 0.01 eV/{\AA}. The two bottom layers, which are expected to be little affected by the
adsorption happening at the surface, were kept fixed in the relaxation
process.

In order to compare the calculations of the adsorbed
molecule-substrate system where periodic boundary conditions were
imposed with those for the isolated molecule in the gas phase, we have
also performed calculations for the molecule placed in a slightly
distorted cubic cell with a cell parameter of 30~{\AA}. This
lengthscale ensures a negligible interaction between the contents of
adjacent cells and mimics the properties of the molecule in the gas
phase. A slightly distorted cell was used to break the symmetry in the
periodic setting.  Apart from the number of k-points (a single k-point
at $\Gamma$ was considered for the gas phase), the same parameters
were employed for (i) the gas phase calculations (ii) the substrate
calculations and (iii) the molecule with substrate calculations.

Van der Waals corrections -also called dispersion corrections- which
originate from long-range electron interactions are not captured by
the standard DFT. Several approaches have been developed to improve
this
situation~\cite{jpcc-2009-113-10541,prl-2009-102-073005,prl-2011-107-245501,mrsb-2010-35-435}.
In our calculations, we consider these effects by using an empirical
DFT energy correction scheme (DFT-D) developed by Grimme and
implemented in VASP~\cite{jcc-2006-27-1787}. This implementation has
been demonstrated to be very effective for the description of
molecular systems such as small molecular adducts, $\pi$-stacking, and
large complexes where dispersion corrections are
relevant~\cite{pccp-2006-8-5287,obc-2007-5-741,jcc-2007-28-555}.

The dispersion corrected DFT-D energy is calculated by adding an
empirical correction $E_{\rm Disp}$~\cite{jcc-2006-27-1787} to the
Kohn-Sham energy:
\begin{equation}
E_{\rm DFT-D} = E_{\rm KS-DFT} +E_{\rm Disp}
\end{equation}
\begin{equation}
 E_{\rm Disp} = -s_6 \sum_{i=1}^{N-1} \sum_{j=i+1}^{N} \frac{C^{ij}_6}{R_{ij}^6} f_{\rm dmp}(R_{ij}) 
\end{equation}
$C^{ij}_6$ is a dispersion coefficient for the atom pair $ij$ which is calculated by:
\begin{equation}
C^{ij}_6 = \sqrt{C^{i}_6 C^{j}_6 }
\end{equation}
where the coefficient $C_6^i$ for atom $i$ is given by
$C^i_6$=0.05$NI^i_{\rm p}\alpha^i$, $N$ has values 2, 10, 18, 36, and 54 for
atoms from rows one to five of the periodic table, $I_{\rm p}$ is the atomic
ionization potential and $\alpha$ are static dipole
polarizabilities.
\\
$R_{ij}$ denotes the interatomic distance between atoms $i$ and $j$
and $s_6$ is the global scaling factor for the functional. The term
$f_{\rm dmp}(R_{ij})$ is a damping function which is given by:
\begin{equation}
f_{\rm dmp}(R_{ij})=\frac{1}{1+e^{-d\big(\frac{R_{ij}}{R_r}-1\big)}}
\end{equation}
The role of the damping function is to scale the force field such as
to minimize contributions from interactions within typical bonding
distances.  Here, $R_r$ is the sum of atomic vdW radii.  Original
values of all atomic and global parameters proposed by Grimme are
employed in the present study~\cite{jcc-2006-27-1787}. The summations
are performed over all $N$ atoms in the reference cell and all
translations of the unit cell within a cutoff radius of 30~{\AA}.  The
interactions over distances longer than this radius are assumed to
give negligible contributions to $E_{\rm Disp}$ and can be ignored.

The relative stability of the adsorption geometries is evaluated with:
\begin{equation}
\Delta E_{\rm ads} = E_{{\rm system(MCPM+SiO_2)}} -E_{\rm molecule(MCPM,\:g)}- E_{{\rm clean\:surface(SiO_2)}}
\end{equation}
where $\Delta E_{\rm ads}$ is the calculated adsorption energy,
$E_{{\rm system(MCPM+SiO}_2{\rm )}}$ corresponds to the total energy
of the relaxed system of precursor molecule adsorbed on the substrate,
$E_{\rm molecule(MCPM,\:g)}$ is the total energy of the isolated
precursor molecule calculated in the gas phase, and $E_{{\rm
    clean\:surface(SiO}_2)}$ is the total energy of the clean
hydroxylated SiO$_2$ substrate.

\section{Results}\label{sec:results}

\subsection{Isolated precursor molecule}

\begin{figure}
\begin{center}
\includegraphics[width=0.7\textwidth]{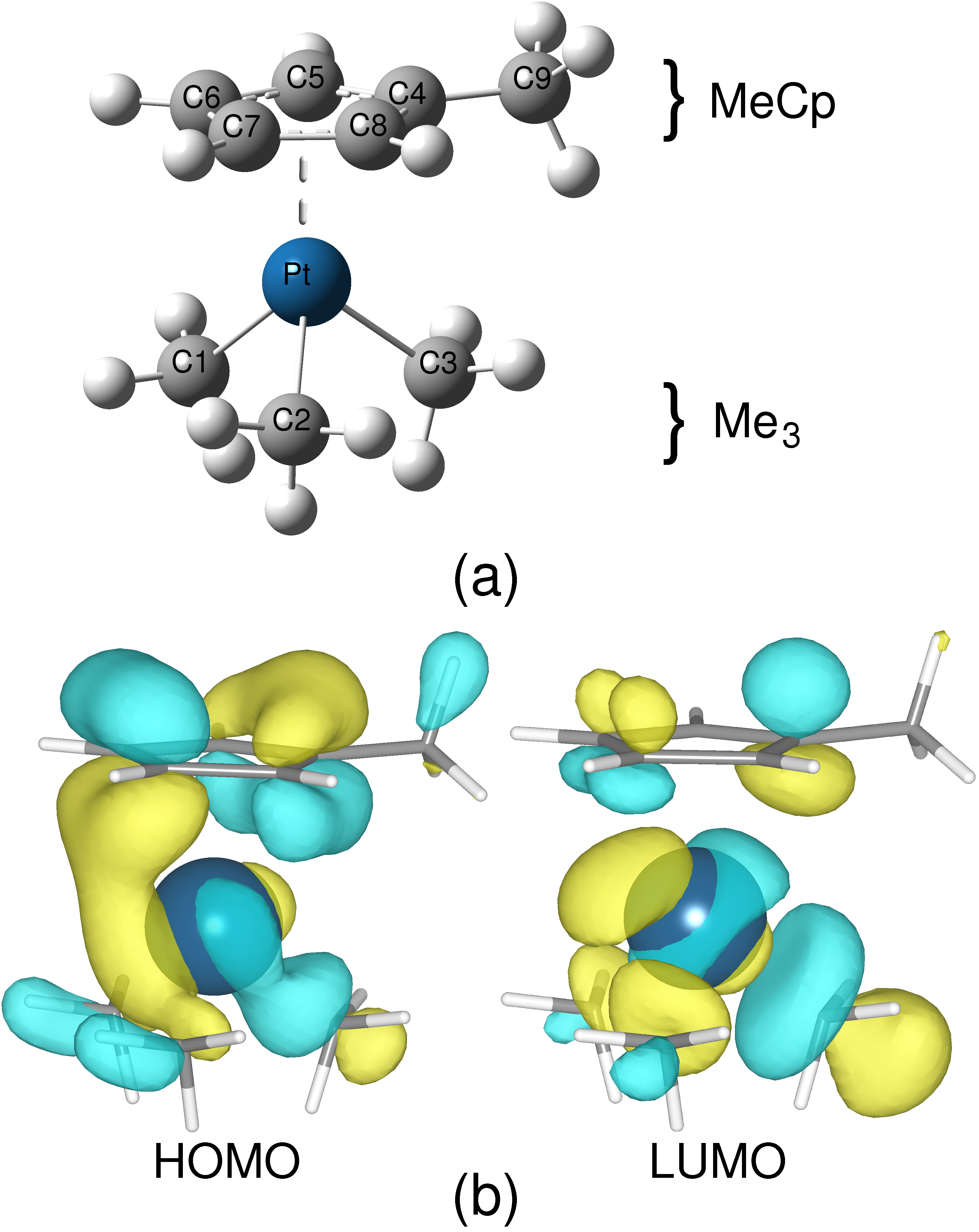}
\end{center}
\caption{(a) Optimized structure of trimethyl methylcyclopentadienyl
  platinum (MeCpPtMe$_3$) in the gas phase with Pt in blue, carbon in
  gray and hydrogen in white. MeCp and Me$_3$ groups are marked. (b)
  Frontier molecular orbitals of MeCpPtMe$_3$. }
\label{fig:molecule}
\end{figure}

Figure~\ref{fig:molecule}~(a) shows the optimized structure of the
MeCpPtMe$_3$ molecule, which has one methyl substituted
cyclopentadienyl ring (MeCp) and three methyl ligands (Me$_3$) bound
to a central Pt atom.  The methylcyclopentadienyl unit is bound to the
platinum centre atom in a pentahapto fashion with the typical
platinum-carbon distances as found in experiment (2.262 to 2.356
{\AA})~\cite{jacs-1989-111-8779}. Table~\ref{tab:bondlengths} shows
selected bond lengths and angles obtained in our calculations compared
with experimental data.  The agreement between both sets of data is
rather good.  After relaxation, the distances between the
cyclopentadienyl ring carbon and the Pt atom show changes of up to
0.11~{\AA} compared to the experimentally measured distances.
The differences for angles between
experiment and theory are mostly within 2$^\circ$.

\begin{table}
  \caption{
    Calculated bond lengths ($d$) and angles ($\varphi$) compared to 
    the corresponding experimental values for MeCpPtMe$_3$. Distances 
    are given in {\AA}, and angles are given in degree. The bond 
    length $d_{\rm C(Cp\:ring)-H}$ is 1.088~{\AA}, $d_{\rm C(Me)-H}$ 
    is about 1.099-1.103~{\AA}, and the bond length 
    $d_{\rm C(Cp\:ring)-C(Cp\:ring)}$ is about 1.423-1.442~{\AA} 
    in the calculation.}
\begin{tabular}{cccccc}
\hline\hline
&$d_{\rm C1-Pt}$&$d_{\rm C2-Pt}$&$d_{\rm C3-Pt}$&$d_{\rm C4-Pt}$&$d_{\rm C5-Pt}$\\\hline
Exp.~\cite{jacs-1989-111-8779}&2.019&2.141&1.990&2.265&2.262\\
Calculated&2.073&2.075&2.068&2.376&2.341\\\hline
&$d_{\rm C6-Pt}$&$d_{\rm C7-Pt}$&$d_{\rm C8-Pt}$&$d_{\rm C4-C9}$&\\\hline
Exp.~\cite{jacs-1989-111-8779}&2.317&2.356&2.327&1.508&\\
Calculated&2.321&2.326&2.346&1.499&\\\hline
&$\varphi_{\rm C1-Pt-C2}$&$\varphi_{\rm C2-Pt-C3}$&$\varphi_{\rm C1-Pt-C3}$&$\varphi_{\rm C5-C4-C9}$&$\varphi_{\rm C8-C4-C9}$\\\hline
Expt.~\cite{jacs-1989-111-8779}&84.1&86.7&85.9&127.9&124.0\\
Calculated&84.7&84.9&86.4&126.1&126.3\\
\hline\hline
\end{tabular}
\label{tab:bondlengths}
\end{table}

In order to obtain a first view of how the charge density distributes
in the molecule, we have calculated the electronic properties of an
isolated molecule with the Gaussian 09 code~\cite{gauss09} by
considering the standard basis sets.  The
structure~\cite{jacs-1989-111-8779} of the isolated precursor molecule
MeCpPtMe$_3$ was first optimized and the resulting structure is
consistent with both experimental values and the above GGA calculation
for a molecule in the gas phase.

In Figure~\ref{fig:molecule} (b) we show the calculated molecular
orbitals for MeCpPtMe$_3$. The highest occupied molecular orbital
(HOMO) is mostly formed by C $2p$-type atomic orbitals from the
cyclopentadienyl ring and Pt $5d$-type atomic orbitals. The lowest
unoccupied molecular orbital (LUMO) is mostly composed of a
combination of $2p$-type atomic orbitals of C bonded to  Pt
$5d$-type atomic orbitals. Such an orbital configuration indicates
that a transition from HOMO to LUMO could happen by charge transfer
from the MeCp group to the Me$_3$ group through Pt.

\subsection{Adsorption of  MeCpPtMe$_3$ on the hydroxylated SiO$_2$ surface}

\begin{figure}
\begin{center}
\includegraphics[width=0.8\textwidth]{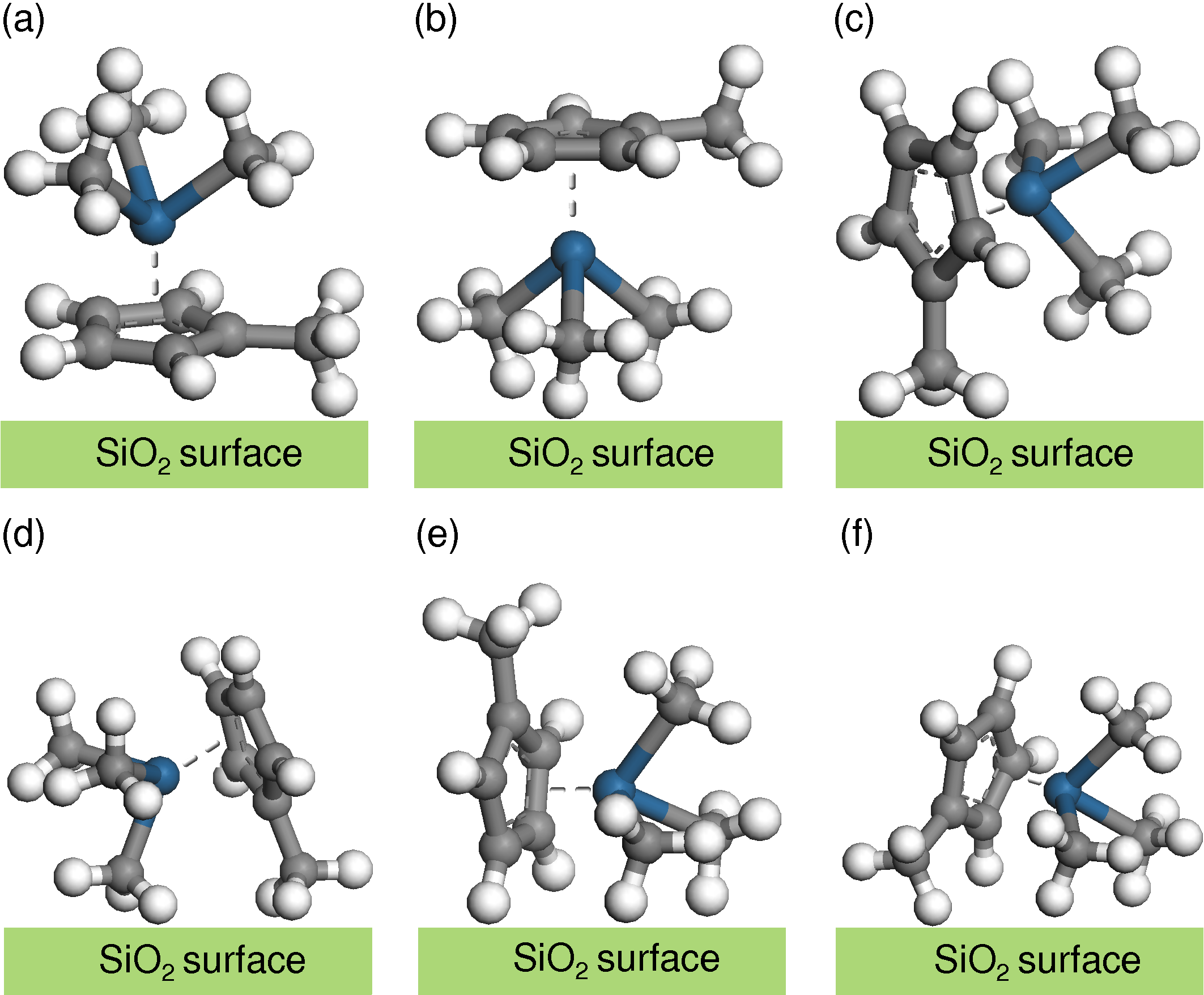}
\end{center}
\caption{The configurations of MeCpPtMe$_3$ contact to the surface
  that were considered in this work (side view). (a) MeCp ring
  attaching, denoted as $O_{\rm MeCp}$; (b) Me$_3$ group attaching,
  denoted as $O_{\rm Me_3}$; (c) Me from MeCp attaching, denoted as
  $O_{\rm Me}$; (d) one methyl from Me$_3$ and one methyl from MeCp
  attaching, denoted as $O_{\rm Me-Me}$; (e) two methyl from Me$_3$ and Cp
  ring attaching, denoted as $O_{\rm Me_2-Cp}$; (f) two methyl from Me$_3$
  and MeCp attaching, denoted as $O_{\rm Me_2-MeCp}$.}
\label{fig:orientation}
\end{figure}

\begin{figure}
\begin{center}
\includegraphics[width=0.8\textwidth]{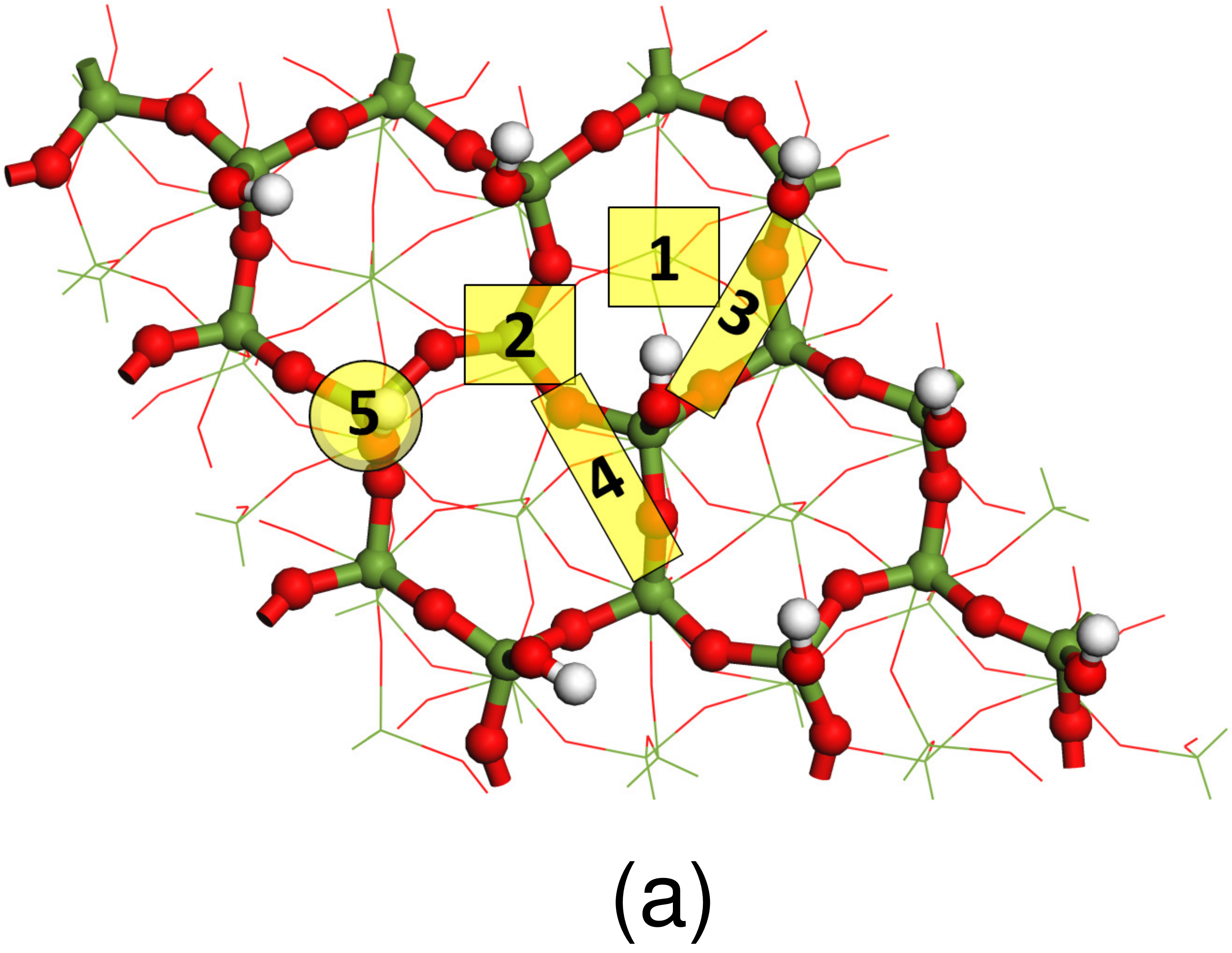}
\includegraphics[width=0.8\textwidth]{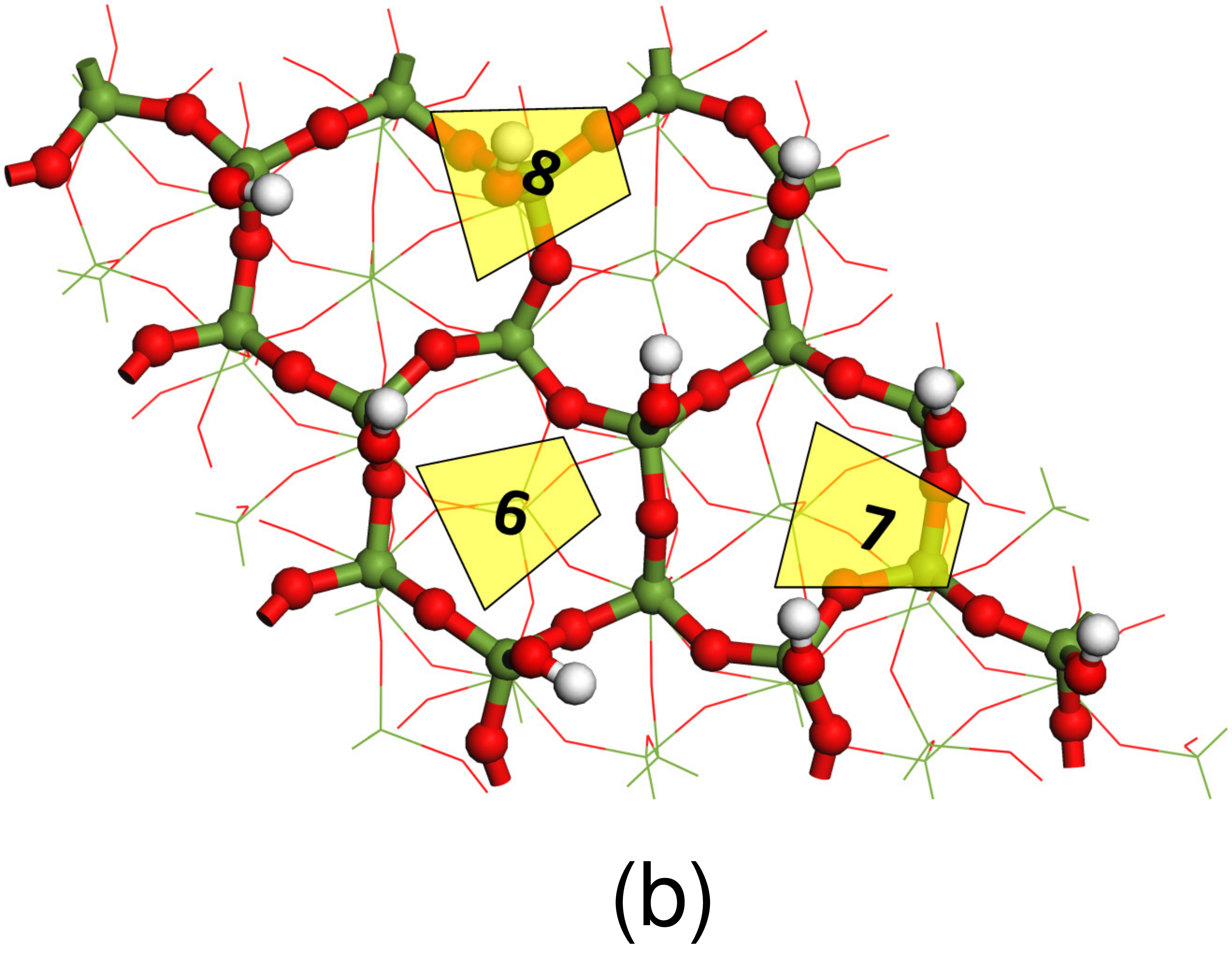}
\end{center}
\caption{Classification of sites in SiO$_2$ (top view) considered
  suitable for adsorption (active sites). (a) Active sites in SiO$_2$
  for one of the ligand groups (MeCp, Me or Me$_3$ groups,
  configurations (a)-(c) in Figure~\ref{fig:orientation}). The sites
  can be classified as hollow sites 1 and 2, bridge sites 3 and 4, and
  Si-OH top site 5; (b) Active sites in SiO$_2$ when both MeCp and
  Me$_3$ groups contact the surface (configuration (d)-(f) in
  Figure~\ref{fig:orientation}), with hollow site 6, near bridge site
  7, and top site 8. The longer border of the trapezoids represents the
  MeCp group and the shorter border represents the Me$_3$ group.  }
\label{fig:site}
\end{figure}

Figure~\ref{fig:sio2} shows the structure of bulk $\beta$-cristobalite
SiO$_2$ and the employed slab model in this study. This slab has
already been relaxed, and the properties analyzed in a previous
work~\cite{muthupaper}.  With this slab, a series of adsorption
geometry models based on the features of the MeCpPtMe$_3$ molecule and
of the SiO$_2$ substrate have been evaluated in order to screen for
the most stable adsorption configuration.  In
Figure~\ref{fig:orientation} we show the various configurations.  If
we consider the molecule as composed of Me$_3$ and MeCp ligand groups
bonding to the central Pt, six different configurations can be
considered.  In three of them only one of the ligand groups faces the
surface (configurations (a)-(c) in Figure~\ref{fig:orientation}) and
in the other three both of the ligand groups face the surface
(configurations (d)-(f) in Figure~\ref{fig:orientation}).  For each
configuration, a set of possible adsorption sites on the substrate are
chosen as shown in Figures~\ref{fig:site} (a) and (b).  For sites 1-5,
the orientation of the molecule on the surface is through either the
MeCp, Me or Me$_3$ groups (see Figure~\ref{fig:molecule} (a) and
Figure~\ref{fig:orientation} (a)-(c)).  For sites 6-8 both MeCp and
Me$_3$ groups can face the surface (configurations (d)-(f) in
Figure~\ref{fig:orientation}).

\begin{table}
\caption{Calculated adsorption energies (in eV) for various adsorption
  configurations without (denoted as ${E_{\rm DFT}}$) and with
  (denoted as ${E_{\rm DFT-D}}$) inclusion of vdW long range
  interaction effects. ${E_{\rm DFT^*}}$ is the DFT energy
  contribution to ${E_{\rm DFT-D}}$. Shown in boldface are the energies of the
most stable configurations for various molecule orientations.}
\vspace{3mm}
\centering
\begin{tabular}{ccccc}
\hline\hline
\multirow{2}{*}{
Orientation (Figure~\ref{fig:orientation})}&
\multirow{2}{*}{
Site (Figure~\ref{fig:site})}&
\multirow{2}{*}{
${E_{\rm DFT}}$}&
\multicolumn{2}{c}{
${E_{\rm DFT-D}}$}\\
\cline{4-5}
&&&${E_{\rm DFT-D}}$&${E_{\rm DFT^*}}$\\\hline
\multirow{5}{*}{$O_{\rm MeCp}$}&1&-0.125&-0.465&-0.138\\
&2&-0.128&{\bf -0.515}&-0.129\\
&3&-0.098&-0.452&-0.073\\
&4&{\bf -0.178}&-0.472&-0.153\\
&5&-0.067&-0.283&0.032\\\hline
\multirow{4}{*}{$O_{\rm Me_3}$}&1&-0.102&-0.394&-0.103\\
&2&-0.109&-0.438&-0.128\\
&3&{\bf -0.173}&-0.446&-0.047\\
&5&-0.068&{\bf -0.496}&-0.030\\\hline
\multirow{2}{*}{$O_{\rm Me}$}&1&-0.077&-0.327&-0.063\\
&2&{\bf -0.111}&{\bf -0.359}&-0.013\\\hline
$O_{\rm Me-Me}$&7&-0.125&{\bf -0.526}&0.148\\\hline
&6&-0.101&-0.477&-0.108\\
$O_{\rm Me_2-Cp}$&7&-0.148&{\bf -0.669}&-0.149\\
&8&{\bf -0.184}&-0.596&-0.098\\\hline
\multirow{3}{*}{$O_{\rm Me_2-MeCp}$}&6&-0.111&-0.503&-0.129\\
&7&{\bf -0.154}&-0.495&-0.089\\
&8&-0.078&{\bf -0.650}&0.000\\
\hline\hline
\end{tabular}
\label{tab:adsE}
\end{table}

\begin{figure}
\begin{center}
\includegraphics[width=0.8\textwidth]{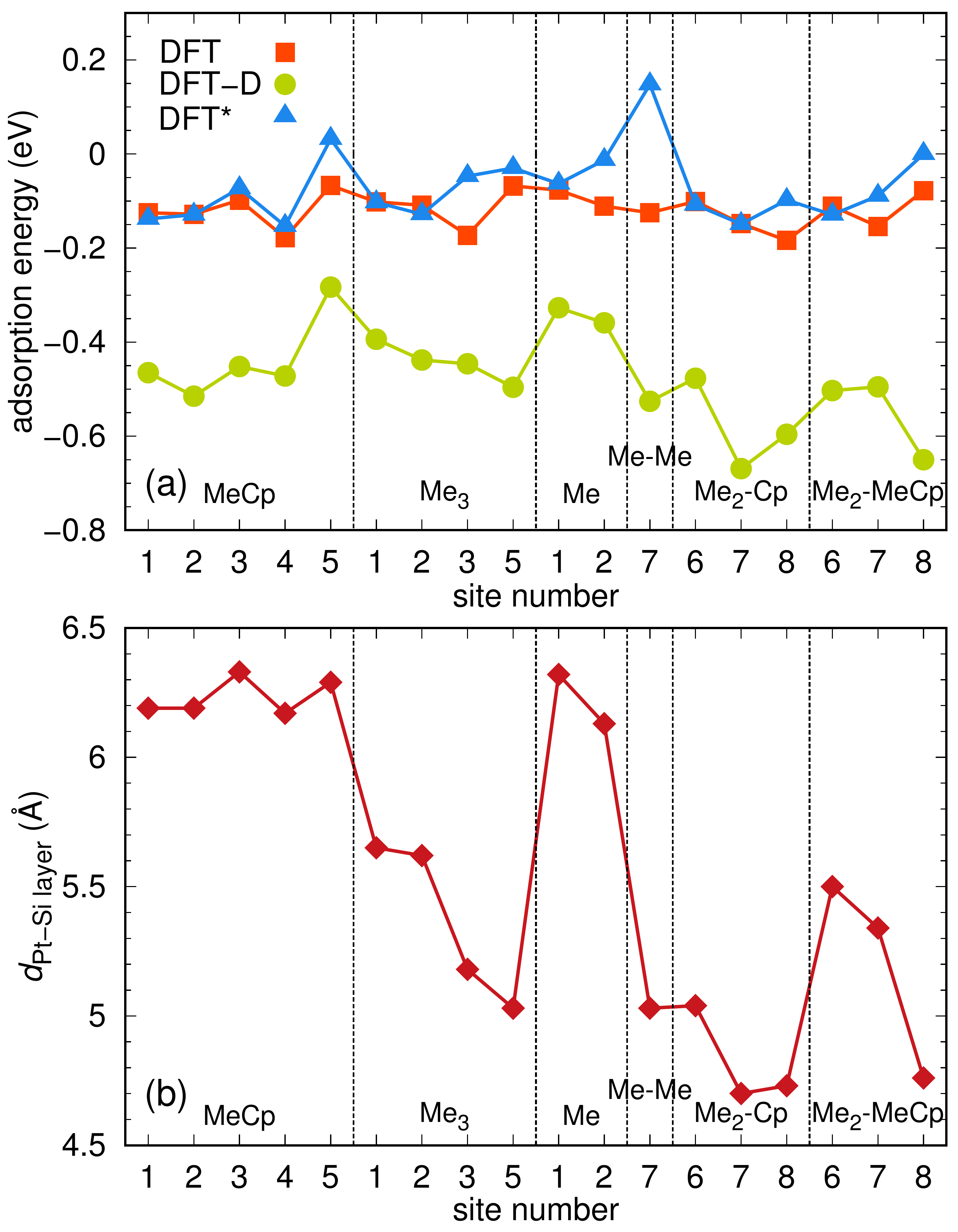}
\end{center}
\caption{(a) Calculated DFT, DFT-D and DFT* adsorption energies; lines
  are a guide to the eyes. (b) Distances between Pt atom and the
  topmost Si plane.  }
\label{fig:adsE}
\end{figure}

\begin{figure}
\begin{center}
\includegraphics[width=0.8\textwidth]{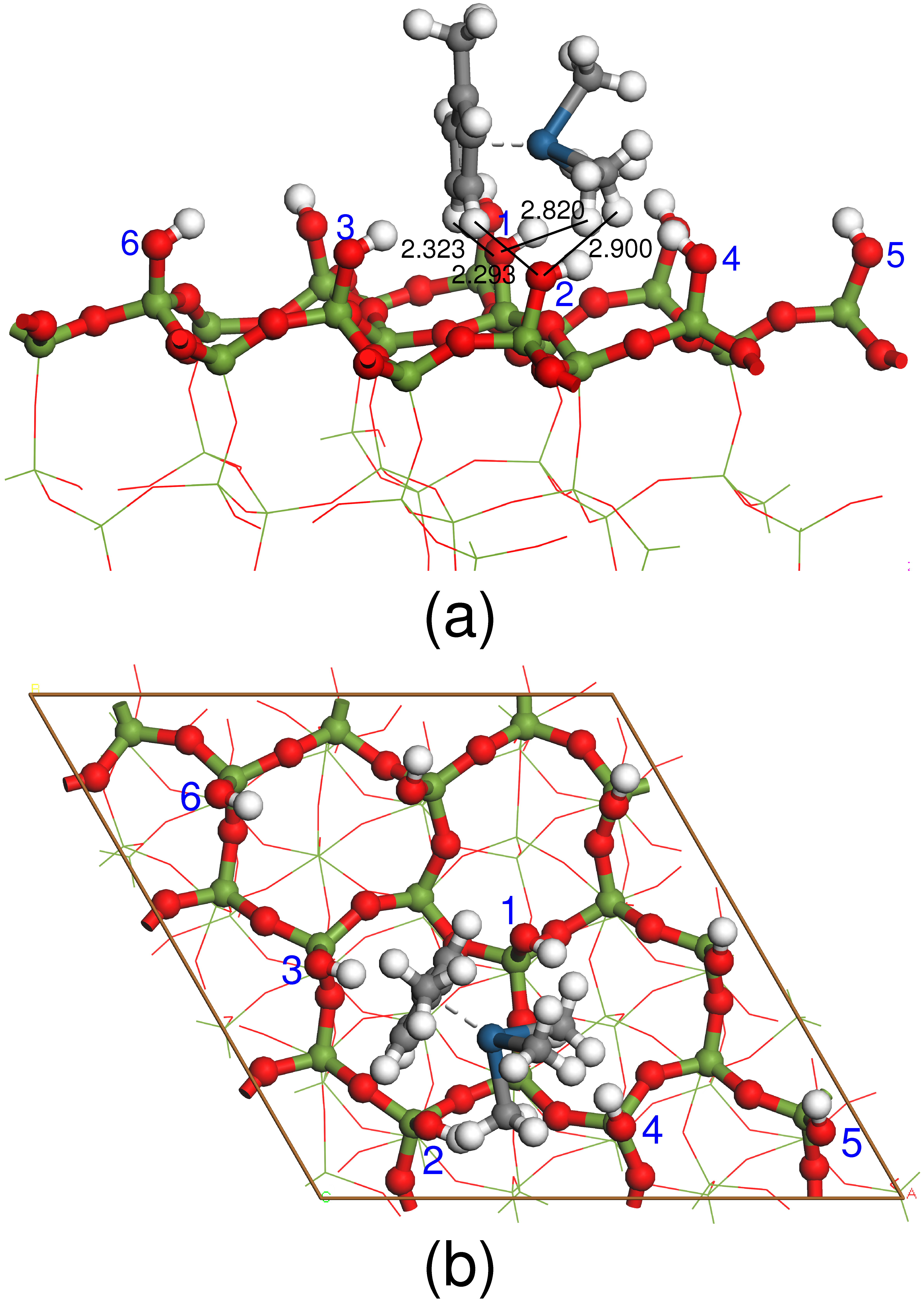}
\end{center}
\caption{The most stable structure found of MeCpPtMe$_3$ adsorbed on
  SiO$_2$ surface, with (a) for side view and (b) for top
  view. Distances in (a) are given in unit of {\AA}, relevant surface
  silanols are numbered in blue.}
\label{fig:moststable}
\end{figure}

Table~\ref{tab:adsE} lists the calculated adsorption energies where
the structure relaxation and the energy calculations were performed
without (denoted as ${E_{\rm DFT}}$) and with inclusion of vdW long
range interaction effects (denoted as ${E_{\rm DFT-D}}$).  We have
also included a third column in the table denoted by ${E_{\rm DFT^*}}$
which contains the DFT energy contributions to ${E_{\rm DFT-D}}$
(i.e. vdW energy corrections have been substracted from ${E_{\rm
    DFT-D}}$).

  In order to analyze these energies, we have plotted
in
Figure~\ref{fig:adsE}  the binding energies of MeCpPtMe$_3$ on
the SiO$_2$ surface as a function of the active adsorption site
without (DFT) and with inclusion of vdW corrections (DFT-D) for all
considered configurations. Also shown is the decomposed DFT
contribution in DFT-D (DFT*).  Within the DFT calculation, the
adsorption site and the orientation of the molecule are found to be
equally important for the adsorption energy.  With inclusion of vdW
corrections, the adsorption energy is primarily determined by the
orientation of the molecule and, to a lesser degree, by the chosen
adsorption position of the molecule on the substrate.  For different
molecular orientations, the DFT-D adsorption energy varies by about
310 meV; meanwhile, for a given orientation, changes in the adsorption
position on the substrate lead to variations in the adsorption energy
between 32 and 232 meV.  The most stable configuration is
significantly modified when vdW corrections are included.  In fact,
the effect of vdW forces to the weak bonding of MeCpPtMe$_3$ to a
fully hydroxylated SiO$_2$ surface is to bring the molecule closer to
the surface. 

If we consider the centered Si from the substrate
 as the topmost Si plane location, the
distance between the Pt atom and this Si plane for different
configurations in the DFT-D calculation is shown in
Figure~\ref{fig:adsE} (b).  As we can see, the difference between
various configurations follows a similar trend as the adsorption
energy calculated in DFT-D (Figure~\ref{fig:adsE}~(a)). The vdW
correction brings the adsorbate molecule closer to the substrate when
both MeCp and Me$_3$ groups face the surface, thus enhancing the
interaction at the interface.  According to Xue {\it et
  al.}~\cite{jacs-1989-111-8779}, it is difficult to see how the
platinum can contact the surface in the ground state of the molecule
so that some dislocation of the $\eta$5-MeCp ring to lower hapticity
is required before the platinum atom of the hydrocarbon complex may
interact with the surface atoms.  In our case, the distance between Pt
and the substrate in the geometry where both Me$_3$ and MeCp groups
face the surface is lower than for other orientations
(Figure~\ref{fig:adsE}), indicating that this configuration enhances
Pt-substrate interaction strength.

As shown in Table~\ref{tab:adsE} and Figure~\ref{fig:adsE},
 the molecule prefers to stay on the
surface with both Me$_3$ and MeCp ligand groups oriented towards the
surface (see Figure~\ref{fig:orientation}~(d)-(f)) while the
orientation with only one of the ligand groups oriented towards the
surface is about 154 to 310 meV less stable (see
Figure~\ref{fig:orientation}~(a)-(c)). Note, that the $O_{\rm Me_2-MeCp}$
configuration on site 8 is only 19~meV higher in energy, making this
structure a close runner-up to the optimal configuration $O_{\rm Me_2-Cp}$
on site 7; in fact both structures are rather similar and the energies
are difficult to distinguish with the precision that we could achieve
in our calculations.

 In Figure~\ref{fig:moststable} we show the most stable 
adsorption configuration 
calculated with inclusion of
vdW corrections found in our work  with the MeCp group oriented
perpendicular to the surface and coupled with one edge of the aromatic
ring to the substrate ($O_{\rm Me_2-Cp}$). After relaxation of the molecule-substrate
complex, we observe no significant structural deformation of the
precursor molecule. The optimized distances between the topmost O atoms
on the substrate and the closest atoms of the MeCpPtMe$_3$ molecule
are shown in Figure~\ref{fig:moststable} (a). The spacing is significantly
larger than typical covalent or hydrogen bond distances. The
structural properties indicate that the precursor molecule is indeed
physically adsorbed on the fully hydroxylated SiO$_2$ surface without
formation of covalent bonds with the substrate.

From the calculated binding energies at different sites, we can
estimate the surface diffusion barriers, which determine the diffusion
ability along different directions~\cite{nl-2009-9-132}.  At low
coverage, the precursor molecule binds to the surface with both Me$_3$
and MeCp ligand groups.  The orientation $O_{\rm Me_2-Cp}$ prefers site 7
while $O_{\rm Me_2-MeCp}$ prefers site 8, with a large thermodynamic
diffusion barrier of 73-155 meV along the [$\bar2$11] direction.  At
high coverage, the molecule can also bind to the surface with only one
of its ligand groups, in the orientations $O_{\rm MeCp}$ and
$O_{\rm Me_3}$, with small diffusion barriers.

After adsorption, the C-C and C-H bonds of the molecule are slightly
weakened, whereas the Pt-C bonds of the Pt-Me$_3$ group are slightly
shortened. The distance between Pt and the MeCp ring are significantly
decreased with the length between Pt and C atoms of the MeCp ring
shortened by 0.068-0.133~{\AA}. The surface region far from the
adsorbate shows almost no changes. However, the O-H bond of the
surface silanol and some Si-O bonds close to the adsorbate are
weakened after adsorption.  This observation indicates that the
precursor molecule is loosely bonded to the substrate.  We found that
Pt $d_{yz}$ and $d_{z^2}$ orbitals and the C $p_y$ orbitals (the
orbital notation is given in the local reference frame of the molecule
where the $xy$ plane is defined parallel to the MeCp group) play an
important role in the interaction between the MeCpPtMe$_3$ adsorbate
and the surface.

The orbital wave function of the HOMO precursor molecule
(Figure~\ref{fig:molecule} (b)) indicates that the $O_{\rm Me_2-Cp}$
orientation has the largest charge density near the substrate orbitals
and can thus lead to more interaction than other orientations. This
explains why this is the most favorable configuration on the
surface. In contrast, the small contribution of the Me from the MeCp
ligand to the HOMO orbital explains why the configuration where this
ligand group faces the substrate corresponds to the most unstable
adsorption configuration (the most unfavorable adsorption energy found
in this work (see Table~\ref{tab:adsE}). 
In previous studies on adsorption of molecules containing conjugated
 $\pi$ electrons in aromatic rings like cyclopentene (c-C$_5$H$_8$) on
 Ni (111)~\cite {jmcaC-2009-314-28}, cyclopentadienyl anion
 (c-C$_5$H$_5$$^-$) on Ni (111)~\cite{ass-2008-254-5831} or pyridine
 on Cu (110) and Ag (110) surfaces~\cite{prb-2008-78-045411}, the
 geometric orientation in equilibrium was found to be the one where
 the cyclic hydrocarbons orient nearly parallel to the surface. In
 those cases the $\pi$ systems contribute to the interaction
 between adsorbate and substrate.  This is not the situation in the
 present more complicated MeCpPtMe$_3$ precursor molecule.  In our
 case, the $p_z$ states of C from the Cp ring are located at about -5
 eV deep down in the valence band.

\begin{figure}
\begin{center}
\includegraphics[width=0.6\textwidth]{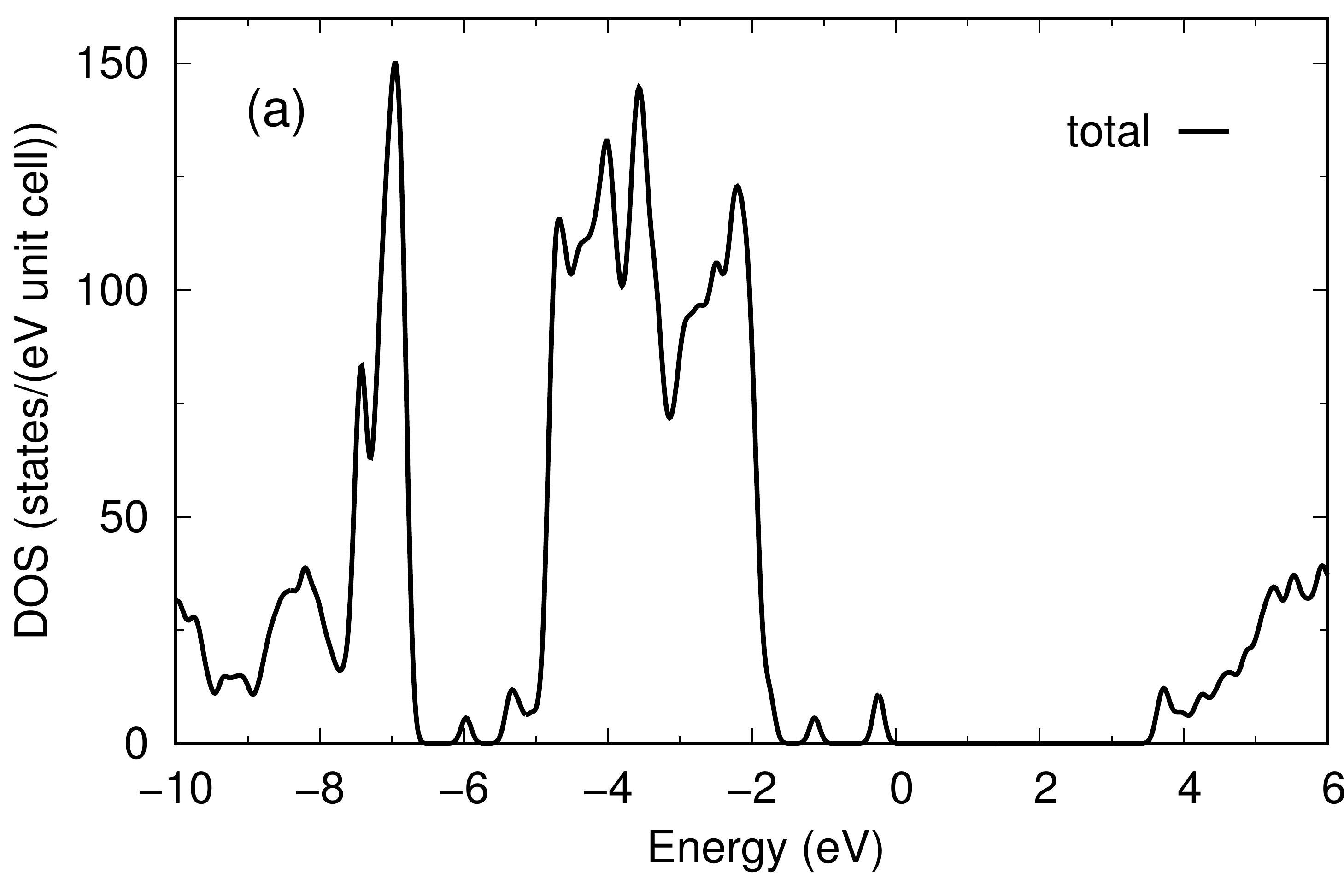}
\includegraphics[width=0.6\textwidth]{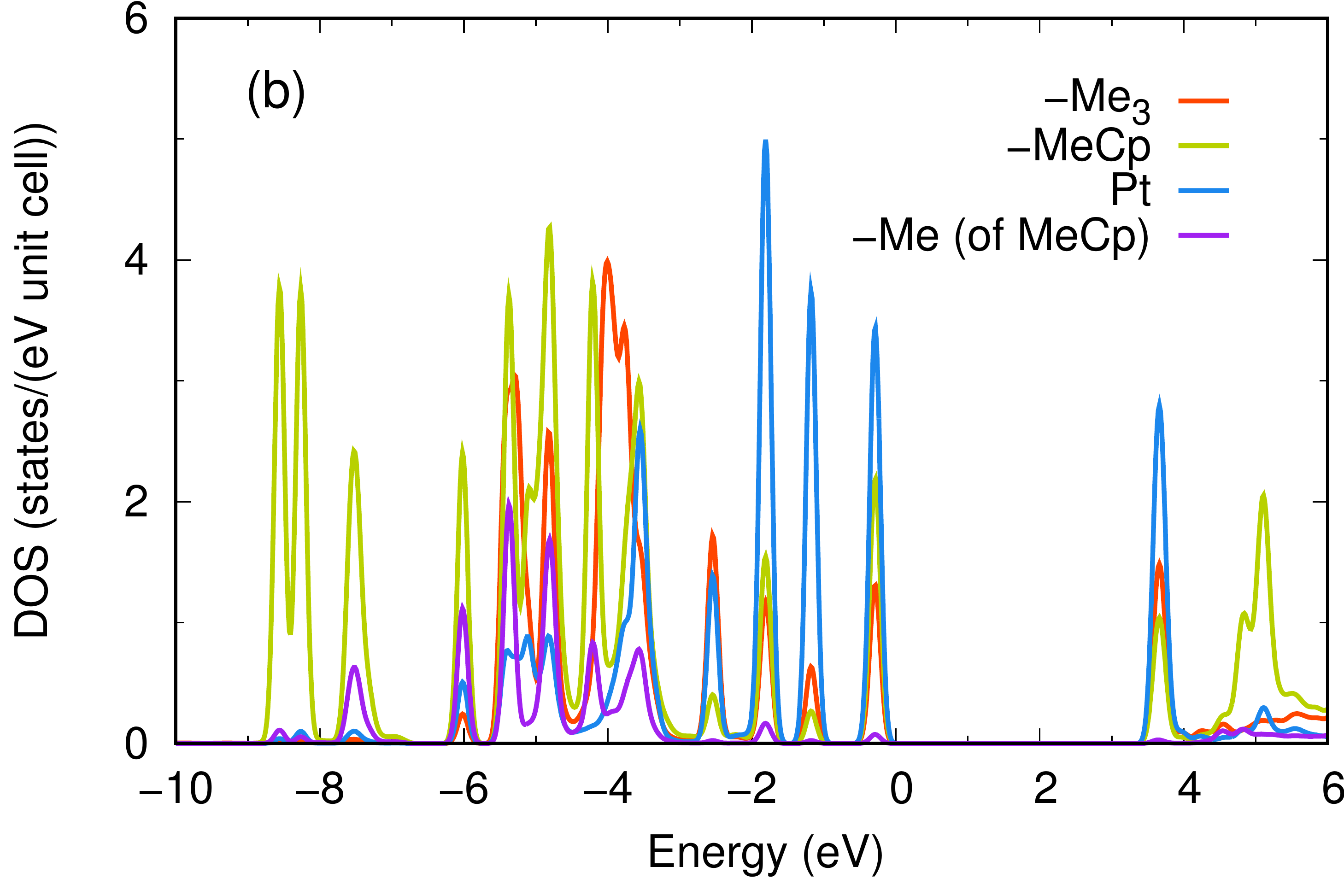}
\end{center}
\caption{ DOS of the preferential configuration of MeCpPtMe$_3$ on SiO$_2$. Shown are 
(a) the total DOS 
and (b) the partial contributions of the adsorbate. The Fermi energy is set to zero.  }
\label{fig:moldos}
\end{figure}

In Fig.~\ref{fig:moldos} we show the total density of states (DOS) of the preferential
 configuration of MeCpPtMe$_3$ on SiO$_2$ as well as the partial contributions of the adsorbate.
 Results are shown
for calculations including vdW corrections. The highest molecular level of the
MeCpPtMe$_3$  adsorbate lies in the gap of the substrate, which pins
the Fermi level close to the valence band.  The largest contribution at the
Fermi level  are Pt and  MeCp states. We observe that the DOS of
the SiO$_2$ substrate shows only minor modifications with respect to the DOS of the free  SiO$_2$
substrate  (compare with Fig. 6 (a) from Ref. ~\cite{muthupaper}).
 These results confirm the weak interaction
between the adsorbate molecule and the SiO$_2$ substrate.

 We can now relate our results to some deposition processes like EBID.
 Electron irradiation of adsorbed MeCpPtMe$_3$ produces hydrogen and
 methane as the only volatile products~\cite{jpcc-2009-113-2487}, and
 the methane could be associated with cleavage of Pt-CH$_3$ (methyl
 from Me$_3$) or C-CH$_3$ (methyl from MeCp).  Figure~\ref{fig:moldos}
 shows that hydrocarbons that contribute to the valence band states
 near the Fermi level are mainly located in the Me$_3$ group and the
 Cp ring of the molecule.  This indicates that the methyl on the
 Me$_3$ group is more active than that of the MeCp group, therefore
 theoretically they will be easier to dissociate when external energy
 is applied.  This is in agreement with recent experimental work where
 the same precursor molecule decomposition on a gold substrate in an
 EBID process has been investigated with X-ray photoelectron
 spectroscopy (XPS) and mass spectrometry
 (MS)~\cite{jpcc-2009-113-2487}. The authors of
 Ref.~\cite{jpcc-2009-113-2487} find that
 trimethyl-methylcyclopentadienyl platinum adsorbed on Au substrates
 at 180 K undergoes electron stimulated decomposition mediated by
 Pt-CH$_3$ bond cleavage instead of C-CH$_3$.

\section{Summary}\label{sec:summary}
Our calculations based on DFT show that the MeCpPtMe$_3$ molecule is
weakly bonded to the fully hydroxylated SiO$_2$ surface without
deformation. The calculated adsorption energy is found to be -0.669 eV
where the molecule is located at a nearly hollow site, and the
distance of Pt to the topmost -OH surface silanol group is 2.87
{\AA}. Contrary to the adsorption of molecules containing conjugated
$\pi$ electrons in aromatic rings, the preferential configuration of
MeCpPtMe$_3$ corresponds to having the MeCp and Me$_3$ groups oriented
towards the surface. We find that the adsorption is more dependent on
the orientation of the adsorbate with respect to the surface than the
adsorption site. We also observe that the optimal configuration of the
molecule corresponds to a minimum in the Pt-substrate distance,
indicating that the interaction between platinum and substrate is
maximized in the lowest energy configuration.  Moreover, inclusion of
vdW corrections contribute to the stabilization of the molecule on the
surface. Our calculations provide a theoretical insight on the
interfaces between MeCpPtMe$_3$ and a SiO$_2$ substrate, which is
essential for the growth processes of deposits.

\section{Acknowledgements}
The authors gratefully acknowledge financial support by
the Beilstein-Institut, Frankfurt/Main, Germany, within the
research collaboration NanoBiC. This work was supported by
the Alliance Program of the Helmholtz Association (Grant
No. HA216/EMMI). The generous allotment of computer
time by CSC-Frankfurt is gratefully
acknowledged.

\end{document}